\documentclass[seceq]{ptptex}

\usepackage{graphicx}

\def \beq{\begin{equation}}
\def \eeq{\end{equation}}
\def\lsim{\raise0.3ex\hbox{$<$\kern-0.75em\raise-1.1ex\hbox{$\sim$}}}
\def\gsim{\raise0.3ex\hbox{$>$\kern-0.75em\raise-1.1ex\hbox{$\sim$}}}

\title{
Net-baryon number fluctuations in (2+1)-flavor QCD%
}

\author{
Christian \textsc{Schmidt}%
}

\inst{
Frankfurt Institute for Advanced Studies, D-60438 Frankfurt am Main, Germany%
}

\abst{
We present a lattice study of net-baryon number fluctuations
 in (2+1)-flavor QCD. The results are based on a Taylor expansion of the 
pressure with respect to the baryon chemical potential. We calculate higher 
moments of the net-baryon number fluctuations and compare with the 
corresponding resonance gas results. We find that for temperature below 
$0.9T_c$ the fluctuations seem to agree with the hadron resonance gas 
predictions. Close to $T_c$, higher moments are increasingly more sensitive 
to the critical behavior of the QCD phase transition. 
Furthermore, we estimate the radius of convergence of the Taylor series as
well as the curvature of the transition line in the temperature chemical 
potential plane.}

\begin{document}

\maketitle

\section{Introduction}
Fluctuations of conserved charges, as baryon number
and strange\-ness are generally considered to be sensitive indicators
for the structure of the thermal medium that is produced in heavy ion
collisions.\cite{Koch} In fact, if at non-vanishing baryon number a
critical point exists in the QCD phase diagram, this will be signaled
by divergent fluctuations of all quantities that can be connected to 
the fluctuations of the chiral order parameter as, e.g., the baryon 
number density.\cite{Stephanov:1998dy,Stephanov:1999zu}

We present here results from lattice calculations of net-baryon number
fluctuations in QCD with dynamical light and strange quark degrees of
freedom.  The results are based on calculations with an improved
staggered fermion action (p4-action) that strongly reduces lattice
cutoff effects at high
temperature and are performed as joined work of the RBC-Bielefeld 
Collaboration.\cite{prep} The values of the quark masses used in this 
calculation
are almost physical; the strange quark mass, $m_s$, is fixed to its
physical value while the light up and down quark masses are taken to
be degenerate and equal to $m_s/10$. This 
corresponds to a pion mass of $m_\pi\approx 220$~MeV.  The lattice
spacing of the lattices we analyze here is with $a\approx 0.25$~fm
($N_\tau=4$) rather large and we expect a cutoff dependence due to a
deformed hadronic spectrum.\cite{deformedHRG}
However, for the quantities we discuss, which are ratios of moments of
net-baryon number fluctuations, the information on details of the
spectrum cancels in HRG model calculations. It thus may be expected
that lattice cutoff effects arising from a distortion of the spectrum
are less severe in such observables. They may be directly comparable to
the hadron resonance gas results and to heavy ion experiments.

\section{The Taylor expansion method}
Direct lattice calculations at nonzero baryon density are impossible
by means of standard Monte Carlo methods.  We follow here the Taylor
expansion approach as described
in detail in Rev.~\citen{our, Allton:2002zi}.  Starting from an expansion
of the logarithm of the QCD partition function, {\it i.e.} the
pressure, one obtains
\beq
\frac{p}{T^4} \equiv \frac{1}{VT^3}\ln Z(V,T,\mu_B)  
=\sum_{i}c_{i}^{B}\left(\frac{\mu_B}{T}\right)^i\; ,
\label{pressure}
\eeq
where
\beq
c_{i}^{B}(T)\equiv\frac{1}{i!}\chi_{i}^{B}
=\frac{1}{i!}\left.\frac{\partial^i}{\partial (\mu_B/T)^i}
\frac{\ln Z(V,T,\mu_B) }{VT^3}\right|_{\mu_B=0}\; .
\label{coeffBS}
\eeq
Here $\mu_B$ is the baryon chemical
potential, which can be obtained as appropriate linear 
combination of the quark chemical potentials. Note that we treat the up 
and down quarks as degenerated, both in mass as well as net-quark number
density. Accordingly, isospin and electric charge chemical potential
vanish. Due to charge conjugation symmetry only even coefficients 
are nonzero. The Taylor expansion coefficients $c_{i}^{B}$ 
are calculated on the lattice as expectation values at $\mu_B=0$. 
Derivatives of the 
partition function with respect to $\mu_B$ are also
know as moments of net-baryon number fluctuations. The second 
derivative defines the quadratic fluctuation $\chi_2^B\equiv 2!c_2^B$, 
while higher derivatives give higher moments ($\chi_4^B, \chi_6^B,\dots$).

\section{The hadron resonance gas}
Within the HRG model interactions are encoded in the thermal creation of 
hadronic resonances, in accordance with their Boltzmann factor.\cite{HRG}
The HRG is known to describe the observed particle abundances in heavy ion
collisions very successfully \cite{HRG_mult} and has recently also shown 
to describe particle fluctuations observed at RHIC.\cite{HRG_fluct}
For the heavy baryons we can utilize Boltzmann approximation and obtain 
for their contribution to pressure
\beq
\frac{p_i^{B}}{T^4}=\frac{d_i}{\pi^2}\left(\frac{m_i}{T}\right)^2
K_2(m_i/T)\cosh(B_i\mu_B/T)\;.\label{HRG_pB}
\eeq 
Here $d_i$, $m_i$ are isospin degeneracy factor and mass of baryon 
species $i$, respectively. 
From Eq.~(\ref{HRG_pB}) we can easily see that the $\mu_B$-dependence 
factorizes and that ratios of moments of baryon number 
fluctuations are either unity ($\chi_4^B/\chi_2^B=\chi_6^B/\chi_2^B=1$) 
or are simply given
by their $\mu_B$-dependence, {\it e.g.} we have 
$\chi_2^B/\chi_1^B=\coth(\mu_B/T)$, 
$\chi_3^B/\chi_2^B=\tanh(\mu_B/T)$.\cite{HRG_fluct}

\section{Results on net-baryon number fluctuations}
\begin{figure}
\begin{center}
\resizebox{0.30\textwidth}{!}{%
  \includegraphics{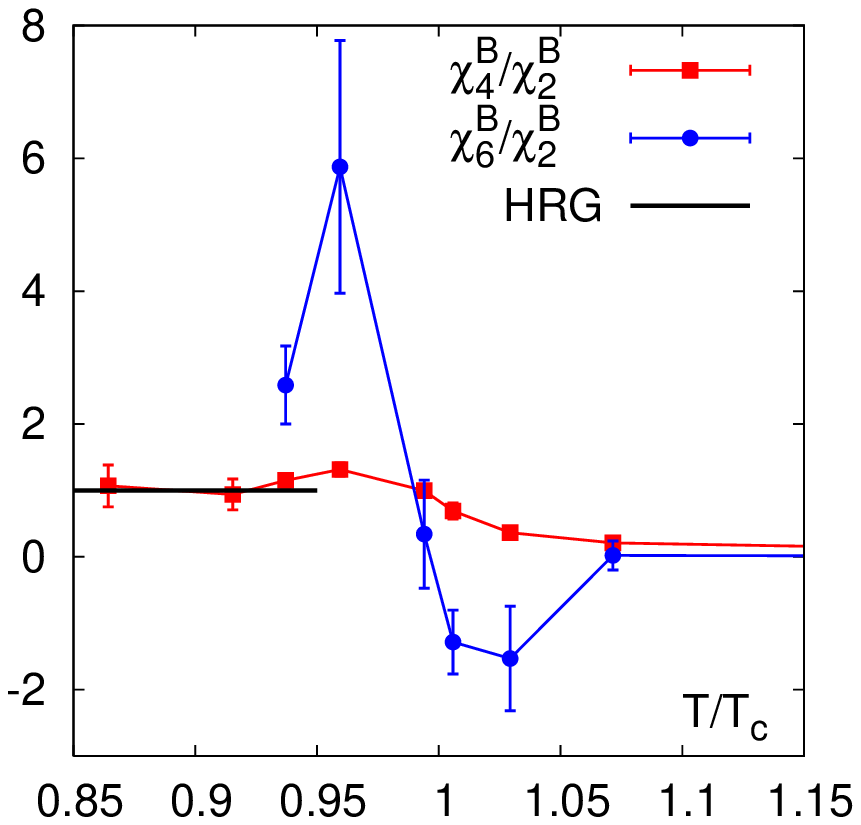}
}
\resizebox{0.30\textwidth}{!}{%
  \includegraphics{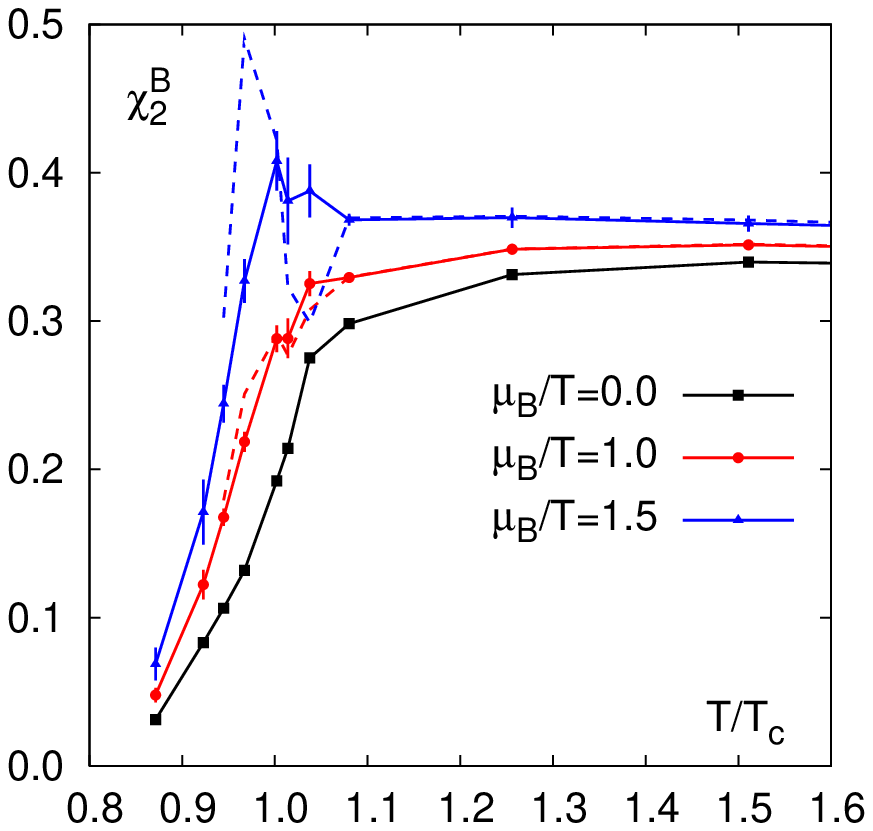}
}
\end{center}
\vspace*{-1mm}
\caption{Shown are $4^{\rm th}$ and $6^{\rm th}$ order net-baryon number 
fluctuations, normalized by the quadratic fluctuations (left) and quadratic 
fluctuations at $\mu_B/T>0$ (right) as a function of temperature.
\label{fig:fluct}}
\end{figure}

\begin{figure}
\begin{center}
\resizebox{0.30\textwidth}{!}{%
  \includegraphics{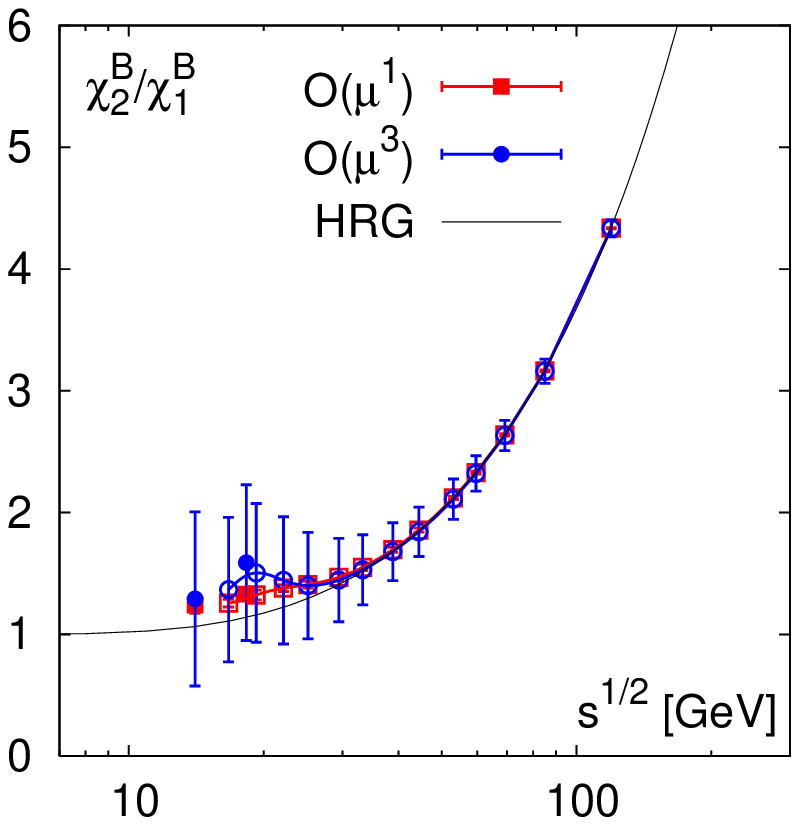}
}
\resizebox{0.30\textwidth}{!}{%
  \includegraphics{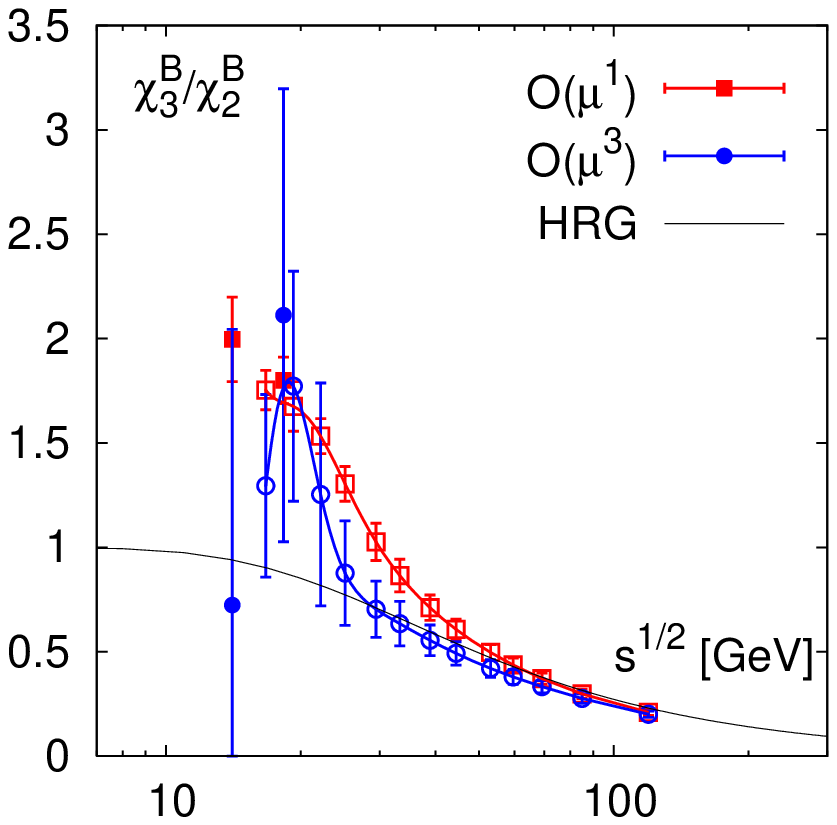}
}
\resizebox{0.30\textwidth}{!}{%
  \includegraphics{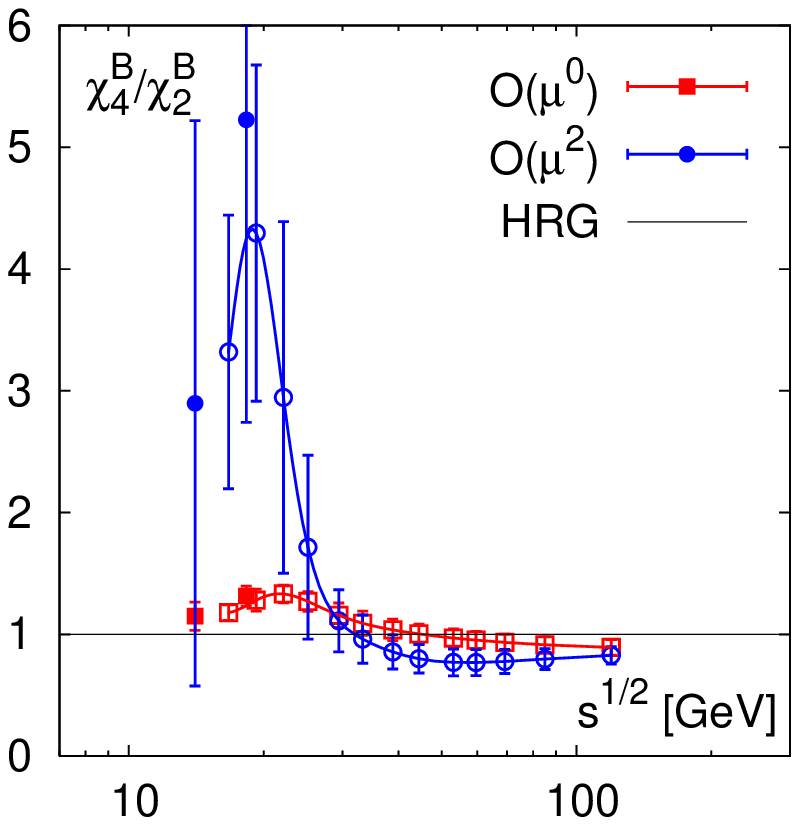}
}
\end{center}
\vspace*{-1mm}
\caption{Shown are ratios of moments of net-baryon number fluctuations 
along the freeze-out curve, as a function of the center of mass energy 
$\sqrt{s}$. \label{fig:freezout}}
\end{figure}

In Fig.~\ref{fig:fluct}(left) we show the 
$4^{\rm th}$ and $6^{\rm th}$ order 
net-baryon number fluctuations, $\chi_4^B$ and $\chi_B^6$, 
normalized by the quadratic fluctuations $\chi_2^B$.
For temperatures $T\lsim 0.9T_c$ the fluctuations seem to agree with
the predicted HRG result which is unity for both quantities. For
temperatures $T\gsim 0.9T_c$ the fluctuations start to deviate from
the HRG model. As expected $\chi_6^B$ increase more
rapidly close to $T_c$, since this quantity diverges in the chiral
limit, where as $\chi_4^B$ will develop a kink.\cite{CPOD}
In Fig.~\ref{fig:fluct}(right) we show the quadratic net-baryon
number fluctuations at $\mu_B>0$. We compare the
$4^{\rm th}$ order (data points) and $6^{\rm th}$ order (dashed lines)
results. We find, that errors from the truncation of the series
(\ref{pressure}) become sizable for $\mu_B/T\approx 1.5$. Finally, 
in Fig.~\ref{fig:freezout} we
plot our results as function of the center of mass energy $\sqrt{s}$ 
along the chemical freeze-out curve by using a suitable parameterization.\cite{freezout}
These results can be directly compared to the net-proton number
fluctuation that are measured in heavy ion experiments. Note
that the ratios have been expanded to obtain their leading order
$\mu_B$-dependence. We expect that for center of mass
energies of $\sqrt{s}\lsim 20$~GeV, {\it i.e.} $\mu_B/T\gsim 1.5$,
higher orders become significant. It is thus important for the
understanding of future heavy ion experiments to obtain $\chi_8^B$ to
good accuracy from first principal lattice calculations.

\section{The radius of convergence}
\label{sec:radius}
\begin{figure}[t]
\begin{center}
\resizebox{0.30\textwidth}{!}{%
  \includegraphics{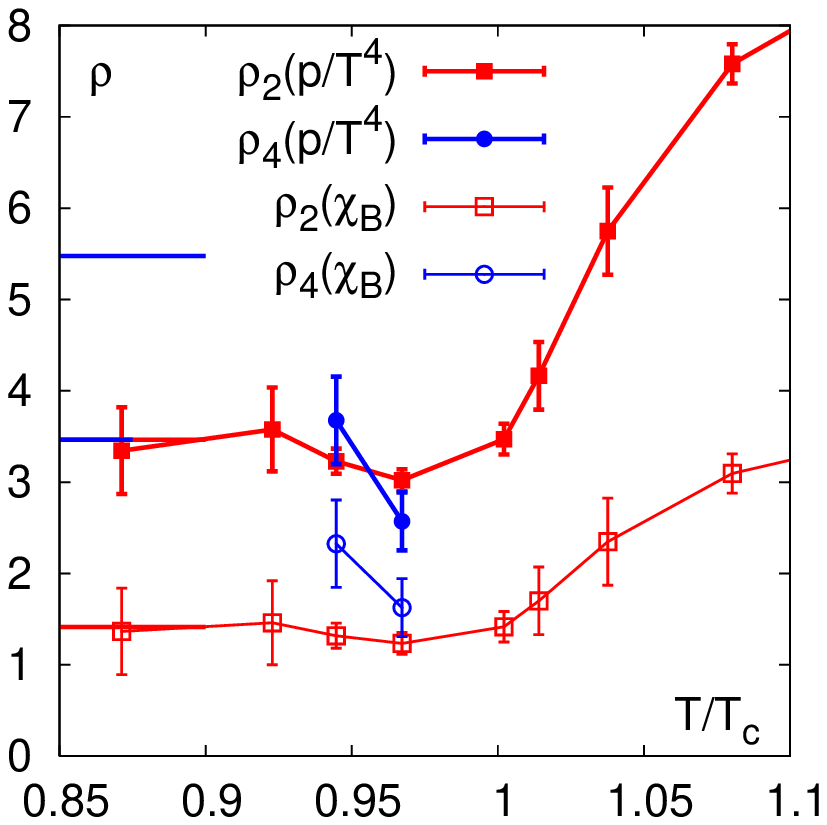}
}
\resizebox{0.30\textwidth}{!}{%
  \includegraphics{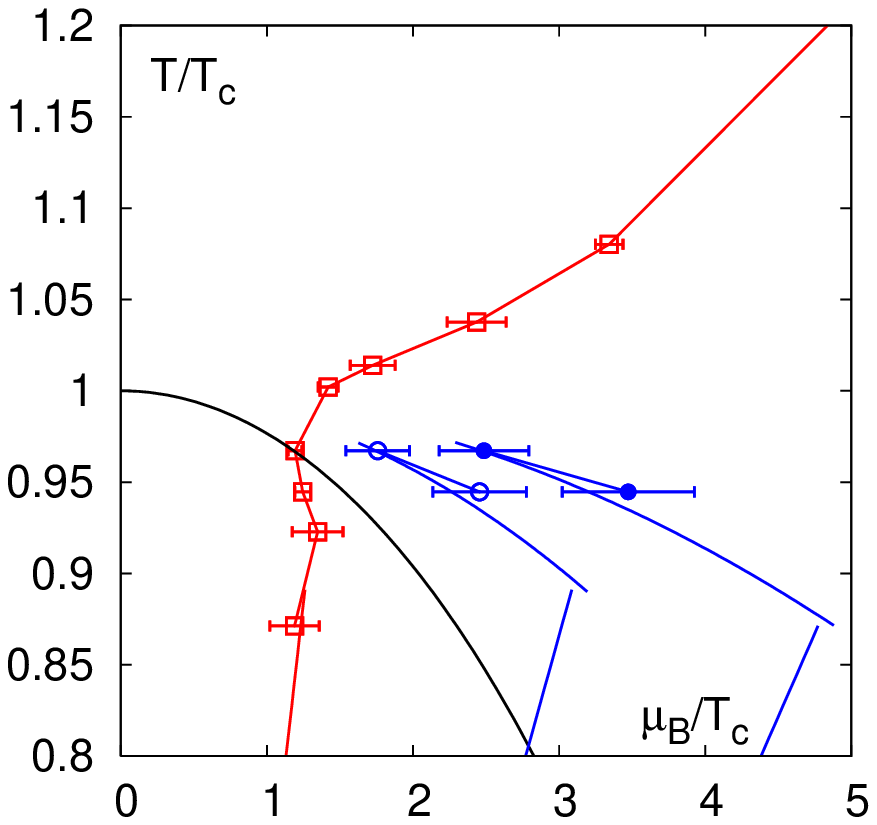}
}
\end{center}
\vspace*{-1mm}
\caption{Estimates of the radius of convergence as explained in the 
text (left), same data plotted in the $(T,\mu)$-diagram, together with 
the freeze-out curve and a band for the transition line 
(right).\label{fig:rho}}
\end{figure}

It has been argued that the radius of convergence can be used to
determine the location of the QCD critical 
point.\cite{CPOD,Allton:2003vx,CEP_GG} One way to define the radius 
of convergence of the pressure series ($\rho[p/T^4]$) 
(Eq.~(\ref{pressure})) is by
\beq
\rho=\lim_{n\to\infty}\rho_n
\quad\mbox{with}\quad
\rho_n=\mu_B^{(n)}/T=\sqrt{c_{n}^{B}/c_{n+2}^{B}}\; .
\label{rho}
\eeq
The radius of convergence can be estimated in a similar manner by the
coefficients in the series of the quadratic net-baryon number
fluctuations ($\rho[\chi_2^B]$). Each order $\rho_n$ will than differ 
by a constant factor which approaches unity in the limit $n\to\infty$. 
In Fig.~\ref{fig:rho}(left) we plot the first two orders of both estimators.
We find, that at $T=0.95T_c$ the estimated radius of convergence is 
consistent with what we find from analyzing the truncation errors, 
{\it i.e.} we find $\rho_4\approx (1.5-2.5) \mu_B/T$.
In Fig.~\ref{fig:rho}(right) we plot the same data in the
($T,\mu_B$)-plane and estimate the curvature of the critical line from
$\rho_4$, assuming that the radius of convergence is limited by the
phase transition. Interestingly the lower limit of the curvature which
we estimate from $\rho_4$ is roughly in agreement with what was found
from an analysis of the O(N) universal scaling.\cite{ON_scaling}

Note that all the results that have been presented here need to be 
confirmed on finer lattices. First $N_\tau=6$ results on $\chi_2^B$ 
and $\chi_4^B$ have already been published.\cite{our}

\section*{Acknowledgements}
Results presented here are based on joint work of the RBC-Bielefeld 
collaboration. Numerical calculations have been performed on the QCDOC 
installations at Brookhaven National Laboratory. This work is partly 
supported by contract DE-AC02-98CH10886 with the U.S. Department 
of Energy and by the Hessian initiative LOEWE through the Helmholtz 
International Center for FAIR.

\end{document}